\documentclass[a4paper]{article}
\input{Qcircuit.tex }
\usepackage{aqis}
\usepackage{graphicx}

\usepackage{amsmath}
\usepackage{float}
\usepackage{caption}
\usepackage{subcaption}
\usepackage{subfloat}

\usepackage{epstopdf}

\begin{document}

\title{
Quantum Circuit Design of Integer Division Optimizing Ancillary Qubits  and T-Count
}

\author{
  Himanshu Thapliyal
  \affiliation{1}
    \email{hthapliyal@uky.edu}
  \and
  T. S.S. Varun
   \affiliation{1}
  \and
  Edgard Munoz-Coreas 
   \affiliation{1}
   }
  
\address{1}{
  University of Kentucky, Lexington, KY, USA
}

\abstract{
In this paper, we present Clifford+T gates based quantum circuit design of  integer division having $n$ ancillary qubits. The proposed quantum circuit is based on restoring division algorithm. The proposed quantum circuit of integer division consists of (i) quantum circuitry of conditional addition  operation, (ii) quantum circuitry of integer subtraction. To design ancillary and T-count optimized design of quantum integer division, the optimized quantum circuit design of integer conditional addition operation and integer subtraction are presented. The proposed quantum integer division circuitry has 50\% improvement in terms of ancillary qubits, and 90\% improvement in terms of T-count compared to the existing design of integer quantum division based on quantum fourier transform. 
}


\maketitle


\section{Introduction and Background}
Quantum circuits of arithmetic operations are vital in designing quantum hardware for Shor's factoring algorithm, solving discrete log problem and quantum cryptanalysis, securing cryptosystems, and circuit design of quantum algorithms such as class number and triangle finding algorithms. 
Dividers are one of the major computational units in quantum arithmetic. Integer division has applications in circuit designs of quantum algorithms, computation of power series, trignometric functions \cite{li2012efficient,yan2013quantum,nielsen2010quantum}. 

Quantum computers of many qubits are extremely difficult to realize; thus, the number of qubits in the quantum circuits need to be minimized. The fabrication constraint of realizing quantum circuits with a large number of qubits has the objective of optimizing the number of ancilla qubits in a quantum circuits.   Designing a scalable and reliable quantum computer is needed now as well as in the future; hence, fault-tolerant quantum circuits are being explored based on Clifford and T gates. "Clifford+T" gate family is illustrated in \cite{amy2013meet, miller2014mapping}.  

In the existing literature, there are a handful of  integer divider designs based on reversible gates targeting mostly reversible computing. Among these designs we found only \cite{khosropour2011quantum} is the only one that is suitable for quantum computing, the usefulness of which in quantum computing is also mentioned in \cite{pavlidis2014fast}. The quantum integer division  in \cite{khosropour2011quantum} uses restoring division algorithm and quantum fourier transform to perform the division operation. However, the design in \cite{khosropour2011quantum} is not optimized for ancillary qubits and T-count. For integer division of size $n$, where $n$ is the number of qubits in the operands, the quantum integer division in \cite{khosropour2011quantum} requires $2n$ number of ancillary qubits, and significant overhead in terms of T-count.  

This paper presents Clifford+T gates based quantum circuit design of  integer division having $n$ ancillary qubits, where $n$ is the number of qubits in the operands. Further, the proposed design has no garbage outputs. The proposed quantum circuit is based on the restoring division algorithm. It employs optimized quantum designs of conditional integer ADD operation, and integer subtraction. Significant improvement in terms of ancillary qubits and T-count compared to the existing design in \cite{khosropour2011quantum} is obtained. Analysis of integer quantum division in terms of T-count  is also presented.

\section{Proposed Restoring Division Algorithm for Quantum Circuits}
The proposed restoring division algorithm for quantum circuits is shown in Table \ref{algorithm1}. In Table \ref{algorithm1}, the inputs to be given are: (a) $ (|Q_{[0:n-1]}\rangle$, $n$ qubit register in which the dividend is loaded ; (b) $|D_{[0:n-1]}\rangle$, $n$ qubit register in which the divisor is loaded; (c) $|R_{[0:n-1]}\rangle$, $n$ qubit remainder register which is initiated to 0 at the start. Therefore, for initiating  $|R_{[0:n-1]}\rangle$, we require $n$ number of ancillary qubits. The algorithm has to go through iteration processes. So, from the algorithm, we can see that at the end of $n$ iterations, we get the quotient at $ (|Q_{[0:n-1]}\rangle$ and remainder at $|R_{[0:n-1]}\rangle$. The divisor is retained at the output. The quantum circuits that are required for developing the hardware implementation of the proposed restoring division algorithm are  (i) Leftshift operation circuitry, (ii) $n$ qubit quantum subtractor and (iii)Conditional ADD operation circuitry. We observed that we can eliminate the LeftShift operation circuitry by combining $|R_{[0:n-2]}\rangle$ and $(|Q_{[n-1]}\rangle$ to form an $n$ qubit register which is actually equal to performing an left shift operation. By combining the qubits in this way, we do not have to use a separate left shift operation circuitry. The block diagrams with a brief explanation of Subtractor and conditional ADD operation quantum circuits  will be discussed in the Sections 2.1 and 2.2, respectively.

\begin{figure}[h]
\centering
\includegraphics[scale=0.75]{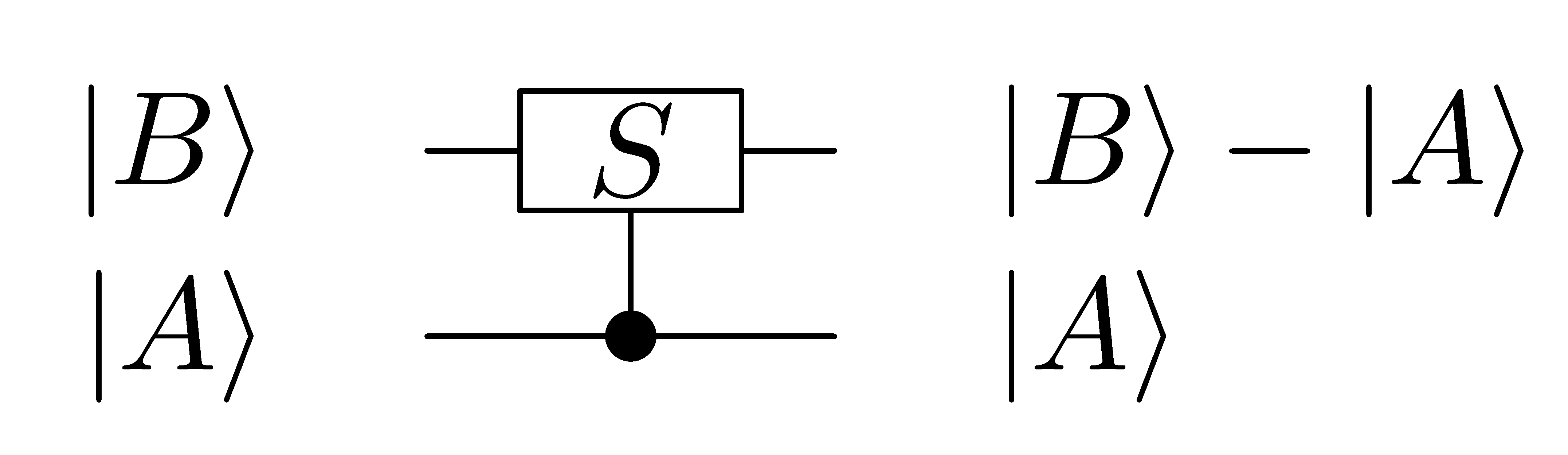}

\caption{Graphic symbol of quantum subtractor. S represents the quantum subtraction operation }
\label{subtractor}
\end{figure}

\begin{figure}[h]
\centering
\includegraphics[scale=0.35]{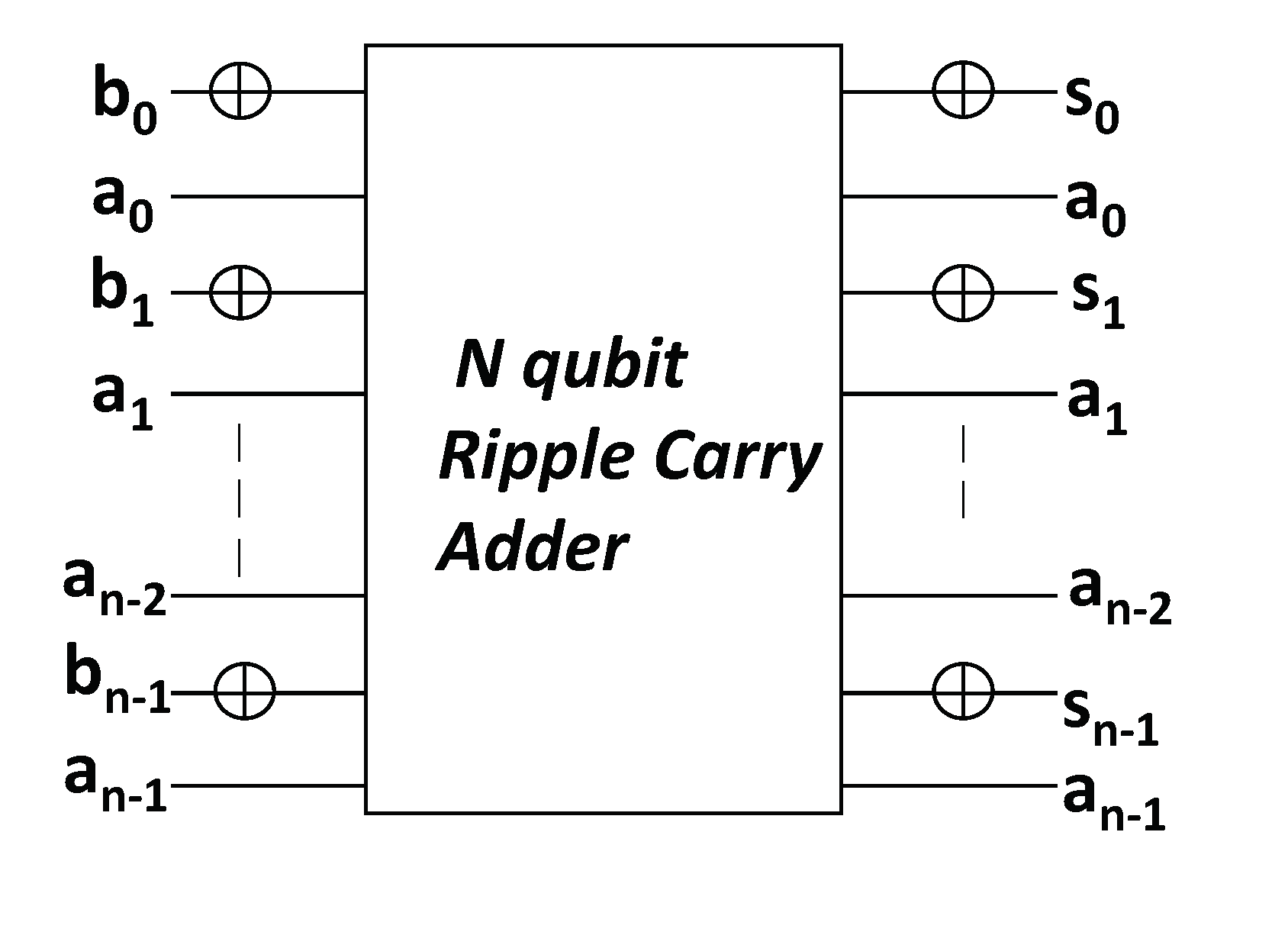}

\caption{Circuit design of N qubits quantum subtractor based on  N qubits quantum ripple carry adder  }
\label{subtractor}
\end{figure}

\begin{figure}[H]
\centering
\includegraphics[scale=0.75]{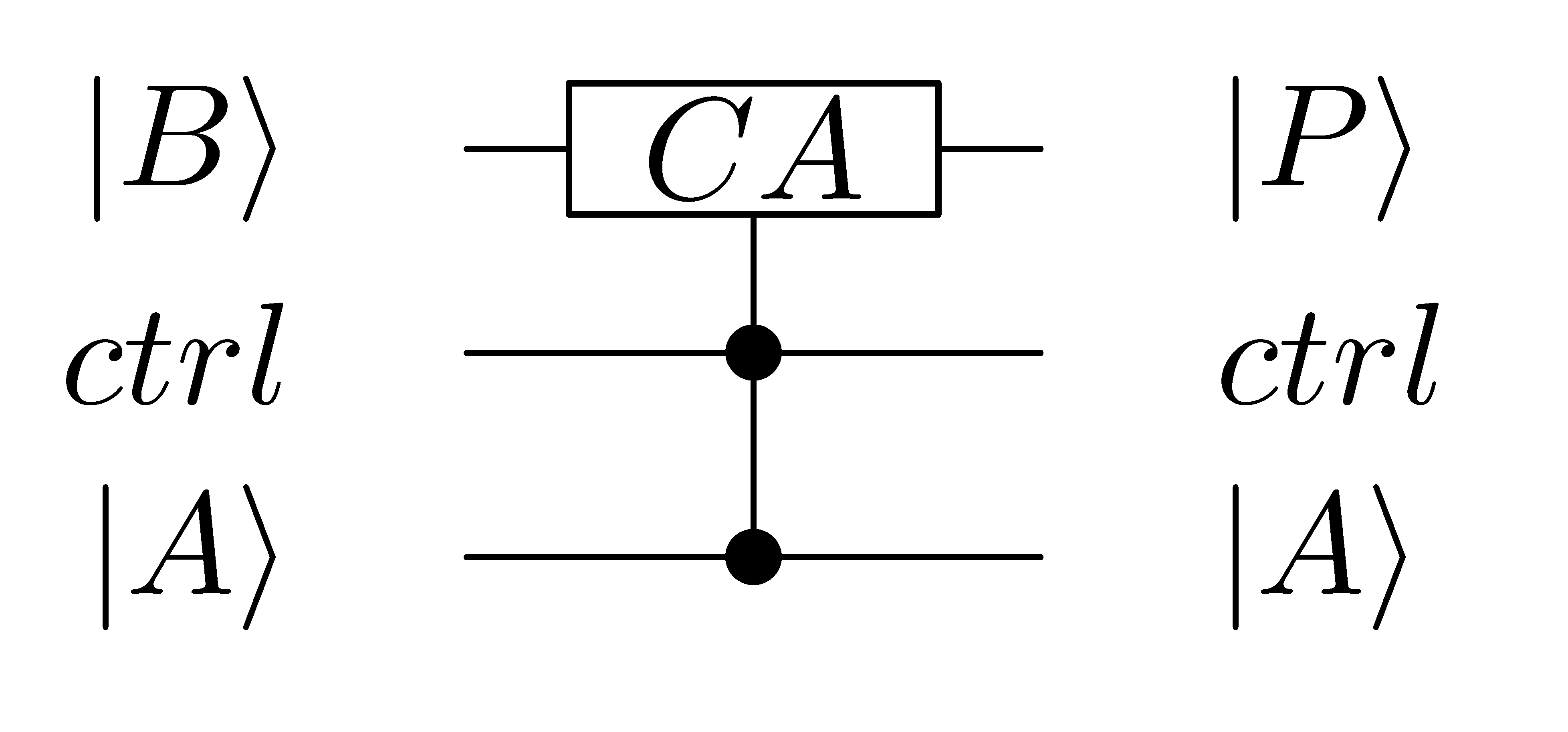} 
\caption{\label{AN block diagram} Graphic symbol of conditional ADD operation circuitry. CA represents the conditional ADD operation.}
\end{figure}

\begin{table}[!h]
\footnotesize
\flushleft
\begin{tabular}{l}

\hrulefill\\

\textbf{Algorithm 1} : \textit{ Proposed Restoring division algorithm }\\

\hrulefill\\

\textbf{function} $Restore\quad(|Q_n\rangle,|R_n\rangle,|D_n\rangle)$\\

\quad\textbf{for} $i=0\quad to\quad n-1$ \textbf{do}\\

\qquad\quad $ (|Q_{[1:n-1]}\rangle,|R_{[0:n-1]}\rangle)$ = L{\tiny EFTSHIFT}  $(|Q_{[0:n-1]}\rangle,|R_{[0:n-1]}\rangle)$;\\

\qquad\quad$(|R-D_{[0:n-1]}\rangle = |R_{[0:n-1]}\rangle-|D_{[0:n-1]}\rangle$;\\

\qquad\textbf{if}$(|R_{[0:n-1]}\rangle > 0 )$ \quad\textbf{then}\\

\qquad\quad $|Q_{[0]}\rangle = 1$\\

\qquad\quad$|R_{[0:n-1]}\rangle = |R-D_{[0:n-1]}\rangle$;\\

\qquad \textbf{else}\\

\qquad\quad $|Q_{[0]}\rangle = 0$;\\

\qquad\quad$|R_{[0:n-1]}\rangle = |R-D_{[0:n-1]}\rangle+|D_{[0:n-1]}\rangle$;\\

\qquad \textbf{end if};\\

\quad\textbf{end for};\\

//repeat for $n$ iterations//\\

\textbf{return} $R$;\\

\textbf{end function}\\

\hrulefill\\

\end{tabular}
\caption{ Proposed Restoring division algorithm	for quantum circuits	}
\label{algorithm1}
\end{table}

\subsection{Design of N Qubits Quantum Subtractor Module}

Fig.\ref{Block dia of subtractor} shows the symbol and brief working of the quantum subtractor  circuitry.  The subtractor circuitry takes two n qubit inputs  $a_{[0:n-1]}$ and $b_{[0:n-1]}$. The input $a$ is regenerated at the output. The n-qubit output $s_{[0:n-1]}$ has the result of the subtraction of $b$ and $a$, i.e., $b-a$. Fig.\ref{subtractor} shows the circuit design of N qubit subtractor based on N qubit quantum ripple carry adder. As shown in Fig.\ref{subtractor},  a quantum ripple carry adder is required to develop a quantum subtractor circuitry. Several examples of quantum ripple carry adder are proposed in the existing literature.\cite{thapliyal2016mapping} \cite{cuccaro2004new} \cite{thapliyal2013design} \cite{takahashi2005linear}. We compared these quantum ripple carry adder circuitries in terms of T-count and ancillary qubits. The comparison results have shown that the quantum ripple carry adder  architecture proposed in  \cite{thapliyal2013design} is superior to the other quantum ripple carry adder circuitries. Hence we used this quantum ripple carry adder for developing the quantum subtractor circuitry. The approach that is followed for developing the quantum subtractor circuitry is $b-a$ = $(\overline{\bar{b}+a})$. Both the inputs are passed through the quantum ripple carry adder. The input qubits $b_{[0:n-1]}$ are complemented at the start and at the end. The qubits $a_{[0:n-1]}$ are just passed through the quantum ripple carry adder.

\subsection{Design of N qubit Quantum Conditional Addition  operation Module }

Fig.\ref{AN block diagram} shows the graphic symbol  of the quantum conditional ADD operation circuit. The  quantum controlled ADD operation circuitry operates as: (i)
when the input labeled $ctrl$ is high (refer Fig.\ref{AN block diagram}), the circuit output is $\ket{P} = \ket{B + A}$, (ii) when the $ctrl$ input is low, the circuit output is $\ket{P} = \ket{B}$.  

\begin{figure}[h]

\centering
\includegraphics[scale=0.75]{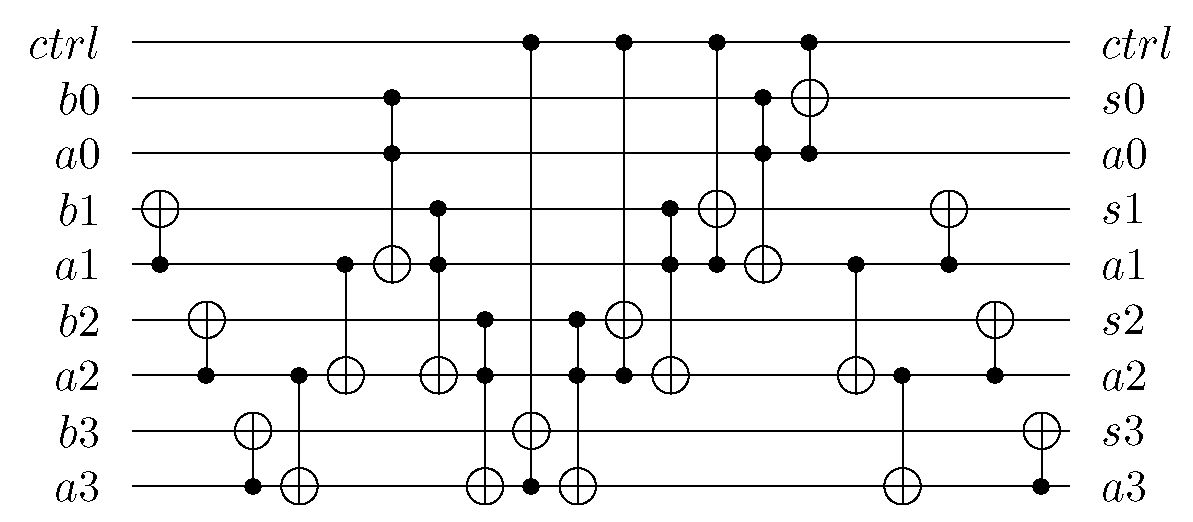}
\caption{Circuit design of quantum conditional ADD operation circuit}
\label{add nop}
\end{figure}

The complete working circuit of quantum conditional ADD operation circuitry is shown in Fig.\ref{add nop} for 4 qubit operands. The quantum conditional ADD circuit uses a modified version of the ripple carry adder proposed in \cite{thapliyal2013design}. We were able to remove the qubit that performs the carry out for the adder in \cite{thapliyal2013design} as we do not need carry out in the computation circuit of quantum conditional ADD. The addition  architecture in \cite{thapliyal2013design} uses Peres gates to perform the addition. The Peres gate can be decomposed into a Feynman and a Toffoli gate. By replacing the Feynman gates with 3 input Toffoli gates, we are able to use the control line ($ctrl$) to perform addition or no operation. Although, Fig.\ref{add nop} is just shown for 4 qubit operands, it can easily be extended to any operands sizes.



\begin{figure}

\centering
\includegraphics[scale=0.75]{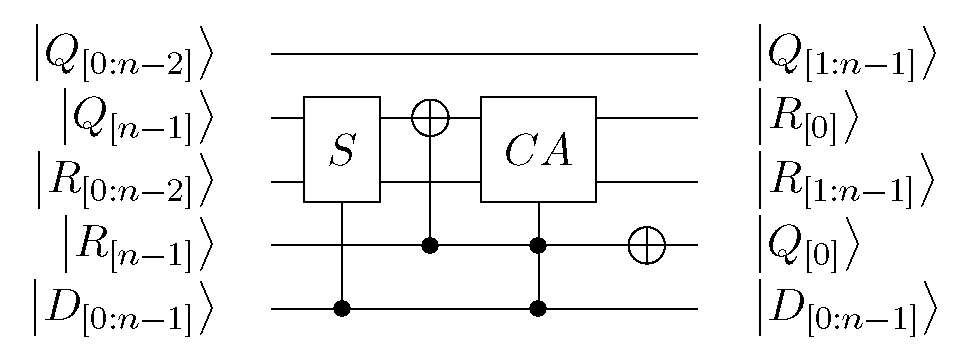}

\caption{Quantum restoring integer divider circuitry design for a single iteration}
\label{restoring 1 iter}
\end{figure}

\begin{figure}

\centering
\includegraphics[scale=0.75]{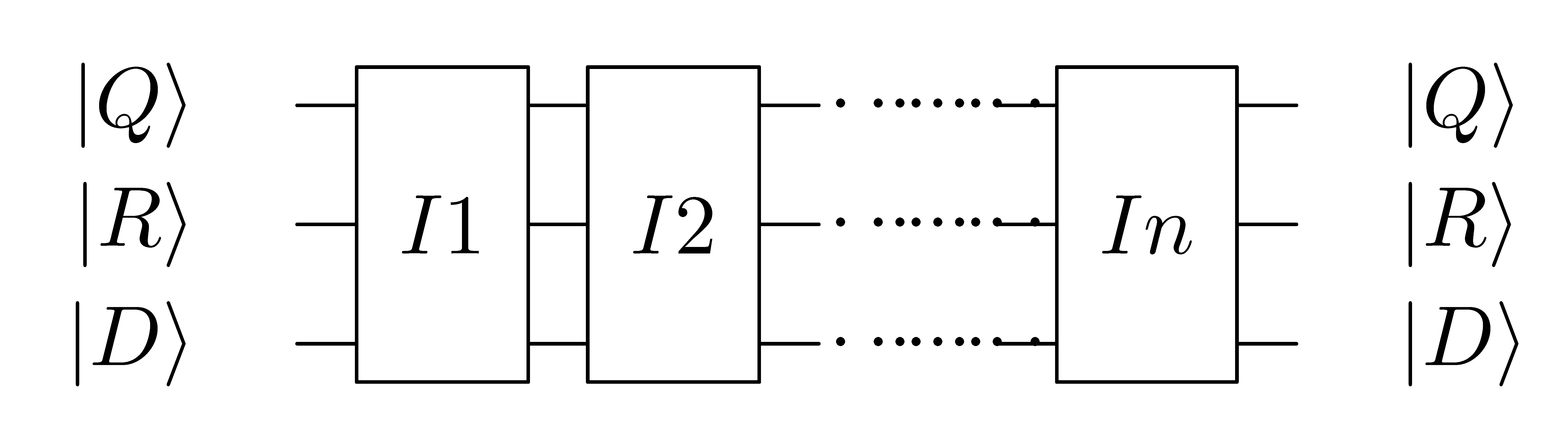}

\caption{Quantum restoring integer divider circuitry design(for $n$ iterations)}
\label{restoring n iter}
\end{figure}




\section{Proposed Design for Quantum Restoring Integer Division  circuitry}

Fig.\ref{restoring 1 iter} shows the proposed quantum circuit of  restoring division for 1 iteration. We now elaborate on how information moves through the circuit.

Step 1. The $ |D_{[0:n-1]}\rangle$ holds the divisor, $ |R_{[0:n-1]}\rangle$ initialised to zero, and $ |Q_{[0:n-1]}\rangle$ holds the dividend. 

Step 2. We consider, $ |Q_{[n-1]}\rangle$ and $ |R_{[0:n-2]}\rangle$, as one combined register. This allows us to not use a left shifting circuit thereby saving quantum resources. 

Step 3. The combined register mentioned above in Step 2, and $ |D_{[0:n-1]}\rangle$ are given as inputs to the quantum subtractor circuitry. Register $ |D_{[0:n-1]}\rangle$ emerges unchanged. The combined register now holds the result of subtraction of $R$ and $D$ registers. Let us call this result as $ |R-D_{[0:n-1]}\rangle$.

Step 4. Qubits $ |R-D_{[n-1]}\rangle$ and $ |R_{[n-1]}\rangle$ are supplied to a CNOT gate. $ |R-D_{[n-1]}\rangle$ is the control qubit and the $ |R_{[n-1]}\rangle$ is the target qubit. The target now holds the value of $ |R-D_{[n-1]}\rangle$ because $ |R_{[n-1]}\rangle$ is always zero throughout the computation.

Step 5. The $ |R_{[n-1]}\rangle$ computed in Step 4 now becomes the control qubit to the conditional ADD circuit. $ |R-D_{[0:n-1]}\rangle$ and $ |D_{[0:n-1]}\rangle$ are the two $n$ qubit inputs to the conditional ADD operation circuit. The outputs of conditional ADD operation are collected. $ |R_{[n-1]}\rangle$ is complemented.

Step 6. All the above operations constitute the first iteration. From the Algorithm in Table 1, we can see that the whole circuit is iterated $n$ times. Hence, the circuit in Fig. 5 is also iterated $n$ times. This is done by using the outputs of first iteration will be used as inputs for the next iteration.

Step 7. This process continues for $n$ iterations. In Fig. \ref{restoring n iter}, $I 1$ represents the first iteration. $I 2$ represents second iteration and $I n$ represents $n^{th}$ iteration. The steps 1 through 6 have to go through each iteration till it reaches $n$ iterations.
This process continues for $n$ iterations.

Step 8. At the end of $n$ iterations, we have Quotient in $ |Q_{[0:n-1]}\rangle$, remainder in $ |R_{[0:n-1]}\rangle$ and the divisor is retained. The dividend is not stored in our implementation.

The resources used in the design of the proposed quantum restoring integer division circuitry is presented in Table \ref{t depth comparision table}. As shown in Table \ref{t depth comparision table}, the proposed design will require $n$ ancillary qubits during initialization of remainder register.  The T-count required by the design is given by summing the cost of subtractor and conditional ADD operation quantum circuitry  at each stage. T-count of the  proposed quantum restoring integer division circuitry is  $35n^2-28n$.

\section{Comparison and Conclusion}
We compared our proposed quantum restoring divider circuitry with the existing design in \cite{khosropour2011quantum}. We compare the ancillaries and T-count.  
The quantum division circuit in \cite{khosropour2011quantum} uses controlled phase shift gates. It is known that realising controlled phase gates other than the controlled T and phase gates with Clifford+T gates cannot be done exactly \cite{kliuchnikov2012fast}. To calculate the T-counts for \cite{khosropour2011quantum} we use T-counts from approximate phase gate implementations reported in \cite{kliuchnikov2012fast}. The implementations with the poorest accuracy were used. This is because the T-count increases significantly as a function of accuracy. Comparison of resource estimation between proposed quantum circuitry of integer division and the existing  quantum circuitry of integer division 
in \cite{khosropour2011quantum} is shown in Table \ref{t count comparision table}. T-count of the existing quantum circuitry of integer division in \cite{khosropour2011quantum} is calculated for 3, 4 and 5 qubits and extrapolated for $n$ qubits. The proposed quantum circuitry of integer division has an improvement ratio of 50\% in terms of ancillary qubits, and 91\% in terms of T-count. 

We presented the general behavioral model of  restoring division algorithm for quantum circuits.  A resource efficient design of the quantum circuitry of integer division is presented by optimizing the quantum circuits modules required by the design, and knitting them together efficiently.  It is observed that non-restoring division algorithm can be an attractive choice to design quantum integer division circuit when minimizing the number of qubits is of primary concern.   The proposed designs can be integrated in a larger data path subsystem designs  to provide resource efficient implementation of quantum algorithms.

\begin{table}[hb]
\caption{Resource count of proposed division circuitry}

\begin{tabular}{|c|c|c|}
\hline

\textbf{ Designs} & \textbf{Ancillaries} & \textbf{T-count} \\
 \hline
 $n$ Subtractor   & 0    & $n*(14n-14)$ \\
 
 \hline
$n$ conditional ADDER  & 0   & $n*(21n-14)$ \\

  \hline
Initial Ancilla qubits& $n$  & 0\\

 \hline
 Total cost & $n$    & $35n^2-28n$ \\
 
 \hline
\end{tabular}

\label{t depth comparision table}
\end{table}

\begin{table}[hb]
\caption{Comparison of resource count between proposed and existing division circuitries }

\begin{tabular}{|c|c|c|c|}
\hline

\textbf{ Designs} & \textbf{Ancillaries} & \textbf{ T-count} \\
 \hline
 existing design \cite{khosropour2011quantum}  & $2n$   &  ~$\approx 400n^2$ \\
 
 \hline
 proposed design  & $n$  & $35n^2-28n$ \\
 
 \hline
 Improvement ratio  & $50\%$  & $\approx 91\%$ \\
 
 \hline
\end{tabular}

\label{t count comparision table}
\end{table}

\bibliographystyle{spmpsci}
\bibliography{instructions}

\bibliography{instructions}

{}

\end{document}